**ARTICLE TYPE**

# USLC: Universal Self-Learning Control via Physical Performance Policy-Optimization Neural Network

Yanhui Zhang[1,2] | Weifang Chen*[1]

[1]School of Aeronautics and Astronautics, Zhejiang University, Hangzhou, P. R. China

[2]Department of Electrical and Computer Engineering, National University of Singapore, Singapore, Singapore

**Correspondence**
Corresponding author is Weifang Chen.
Email: chenwfnudt@163.com

**Abstract**

This study addresses the challenge of achieving real-time Universal Self-Learning Control (USLC) in nonlinear dynamic systems with uncertain models. The proposed control method incorporates a Universal Self-Learning module, which introduces a model-free online executor-evaluator framework to enable controller adaptation in the presence of unknown disturbances. By leveraging a neural network model trained on historical system performance data, the controller can autonomously learn to approximate optimal performance during each learning cycle. Consequently, the controller's structural parameters are incrementally adjusted to achieve a performance threshold comparable to human-level performance. Utilizing nonlinear system stability theory, specifically in the context of three-dimensional manifold space, we demonstrate the stability of USLC in Lipschitz continuous systems. We illustrate the USLC framework numerically with two case studies: a low-order circuit system and a high-order morphing fixed-wing attitude control system. The simulation results verify the effectiveness and universality of the proposed method.

**KEYWORDS**
Universal Self-learning Control, Policy Learning Network, Nonlinear Uncertain Dynamic Systems.

## 1 | INTRODUCTION

With the rapid advancement of aircraft and robotics technology[1], the development of high-performance adaptive control[2] methods for uncertain systems in various scenarios has become a prominent research focus in recent decades[3,4,5]. Reviewing previous works[6,7,8,9,10,11] it is evident that the realization of a multi-task universal adaptive control system with autonomous learning capabilities poses significant challenges.

Adaptive control typically involves parameter matching controllers and adaptive control laws, which dictate online parameter adjustments based on dynamic system responses. Notable theoretical advancements in adaptive control during the 1960s and 1970s emphasized stability and convergence issues, with developments such as Lyapunov Stability Theory[12] and Self-Tuning Regulators[13]. In the 1980s, the maturation of adaptive control theory led to the application of various algorithms for controlling nonlinear uncertain systems, including Gain Scheduling[14], MIT Rule[15], Event-triggered Adaptive Neural[16], Model Reference Adaptive Control[17], and MRAC-Reinforcing Learning[18]. These methods facilitated real-time adjustment of controller parameters without precise knowledge of the system model. The emergence of deep learning as a branch of neural networks in the 1990s provided a novel approach for adaptive learning control[8]. Since the early 21st century, researchers have been enhancing adaptive learning control algorithms and applying them across various engineering domains, such as robotics, aerospace, and intelligent transportation[19,20]. Adaptive learning control demonstrates strong adaptability and efficacy in handling nonlinear or time-varying systems without accurate models.

Lyapunov-type stability arguments and deterministic equivalence principles have been widely employed in constructing stable adaptive feedback strategies for systems with matching uncertainties. However, dealing with unmatched uncertainties in the system has presented significant challenges and prompted the emergence of gradient-based approaches. The Adaptive Control Lyapunov Function (ACLF)[21] was explicitly introduced to address unmatched uncertainties in systems. Nonetheless, computing ACLF necessitates a stabilizing system whose dynamics depend on its own Control Lyapunov Function[22]. Meanwhile, devising autonomous iteration and evolution of controllers based on historical design experience and system data remains a formidable challenge. The primary

**Abbreviations:** Universal Self-learning Control (USLC), Reinforcement Learning (RL)

obstacle in developing such a control framework with self-learning capabilities arises when the object model changes beyond the control input range, rendering the deterministic equivalence principle inapplicable. Consequently, historical controller parameters cannot be directly reused, necessitating offline controller reconstruction.

The aforementioned challenges have sparked a growing interest in data-driven control methodologies. Some of these approaches include reinforcement learning[23], brain-inspired learning control[6], self-learning[24,25,26,27] and lifelong learning[28], large data generated during system control to devise controllers[29]. By circumventing the need for intricate system modeling, these methodologies exhibit applicability across diverse tasks. Moreover, these methodologies are well-suited for real-time control systems necessitating rapid response. Through continuous accumulation and assimilation of real-world data, the data-driven paradigm facilitates iterative enhancement of control policies, rendering it highly adaptable and robust in real-world applications.

Although both adaptive learning control[30] and dazta-driven control[31,32] address the control of uncertain systems, their implementation approaches differ significantly. Adaptive learning control can adapt to various tasks and environments[33]. However, it involves extensive parameter estimation and optimization computations, resulting in high computational complexity that may impair real-time performance. Additionally, adaptive learning[34] is highly sensitive to initialization and parameter selection, and its stability analysis is complex, especially with nonlinear dynamics[35] and parameter estimation, requiring intricate mathematical derivation and extensive simulation validation. In contrast, data-driven methods rely on substantial operational data for learning and optimization, which can lead to overfitting in novel or data-scarce systems. The design and optimization of network models in this approach often depend on empirical judgment and error analysis, making their physical interpretation less transparent.

This paper addresses the challenge of universal self-learning control for nonlinear dynamic systems without precise models. We propose a novel approach that combines adaptive learning with data-driven control based on physical performance data. Specifically, our method involves training a performance policy optimization neural network using historical output data from expert-optimized controllers. This network enables real-time quantification of the iterative optimization policy, allowing for the continuous reconstruction of the controller after each learning cycle. Unlike traditional methods[36] that aim for the final optimization value, our approach focuses on learning the optimization policy, making it applicable to various controllers requiring policy-based optimization. This method is suitable for a wide range of adaptive learning control tasks across different scenarios.

Our main contributions to this paper are as follows: 1) We introduce a novel universal self-learning control framework based on a physical performance optimization policy network, applicable to various universal scenarios. This framework employs the deterministic equivalence principle to design stable adaptive controllers for nonlinear systems with significant uncertainties. 2) We develop a controller policy optimization training dataset using physical performance indicators and construct a neural network to train a universal policy optimization model. This model can handle linear and nonlinear dynamic systems under a time-domain step response evaluation framework. 3) We demonstrate the convergence of the proposed algorithm for a class of nonlinear uncertain systems that satisfy the Lipschitz continuity condition. Specifically, under the conditions outlined in Theorem 1, we prove that for the system described by equation (6), the controller can, after a finite number of online self-learning iterations, find at least one non-empty three-dimensional manifold space where the system achieves stable convergence. This is independent of the initial conditions, ensuring the theoretical stability boundary of the proposed method.

The remainder sections of this article are organized as follows. Section 2 presents the problem statements and preliminaries. Section 3 describes the design framework and algorithmic details of the proposed method. The proofs of stability for the nonlinear system is provided in Section 4, with additional lemmas moved to the Appendices. Section 5 includes numerical simulations of the Universal Self-Learning Control (USLC) about low-order circuit control systems (Section 5.1) and high-order morphing flight control systems (Section 5.2). Finally, present works and suggestions for future research are offered in Section 6.

## 2 | PROBLEM STATEMENT AND PRELIMINARIES

This section derives the problem statements and research object. We presents a performance -optimization learning neural network to verification stability control problem for a class of nonlinear uncertain systems. Unlike traditional methods[37] that estimate model parameters of the uncertain part of the system model[38] or use direct estimation control, our approach iteratively finds optimal controller parameters. Expanded training data through data augmentation with real-world data, akin to the human brain learning process, we transform the challenge of determining optimal controller parameters into learning an iterative optimization policy to achieve optimal performance.

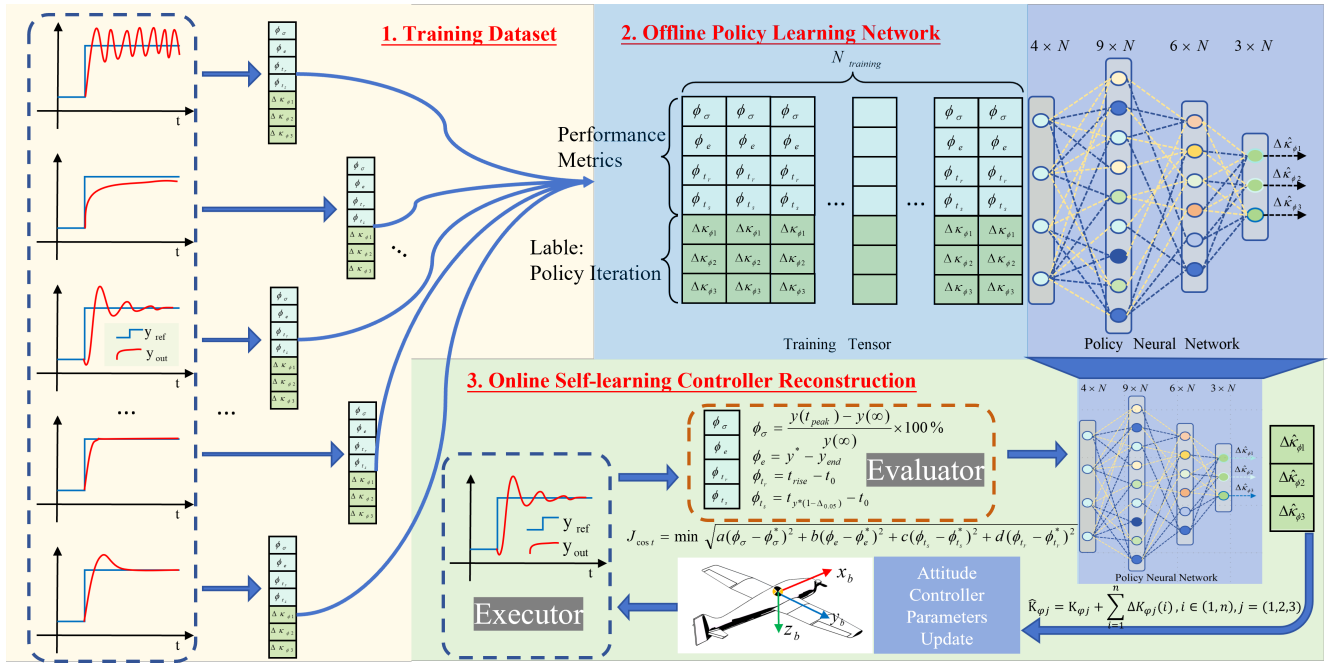

**FIGURE 1** Scheme of USLC: Reconfiguration of online universal adaptive learning controller via optimization neural network based on universal offline pre-trained physical performance policy. The detailed steps are as follows, Step 1: Construct the training dataset based on the physical performance index for the offline optimization policy learning stage, with the data format presented in equation (2); Step 2: Train the policy-optimization learning neural network $\mathcal{N}$ using physical performance indicators and labels from historical human optimization strategies; Step 3: Utilize the real-time output of the policy-optimization model $\mathcal{N}$ to predict the subsequent parameter optimization policy for the current controller. The controller reconstruction rules can be self-learned using the cost function of physical performance. This approach is applicable in general scenarios where controller redesign is required.

**Definition 1.** Consider a class of nonlinear uncertain dynamic systems as follows

$$\begin{cases} \dot{x}_1 = x_2, \quad \text{with} \quad x_1(0), x_2(0) \in \mathbb{R}, \\ \dot{x}_2 = f(x_1, x_2, t) + w * u(t), \\ e(t) = y^* - x_1(t), \end{cases} \quad (1)$$

where $x_1, x_2$ are system state, $e(t)$ is the control error, the system reference command $y^*$ is the ideal controller output under the control law $u(t)$, $w$ is an unknown positive constant, with $\underline{w} > 0$. $f = f(x_1, x_2, t)$ is a continuously differentiable function with upper bounds for state $x_1$ $x_2$, and $f(0,0) = 0$. The control law $u = \alpha(x)$ independent of $\theta$ would have to satisfy $x\dot{x} < 0, x = 0$, and $\theta \in \mathbb{R}$.

*Remark 1.* Although the control objects of different tasks are different, their control objectives are to achieve the best physical performance. Therefore, using the same physical index characteristics as the input of the controller optimization network can ensure the universality of the system optimization neural network (NN). The control goal is to predict the offset $(\Delta \mathcal{K}_{1,2,3})$ of the controller self-learning reconstruction parameters based on the NN model.

*Remark 2.* If the nonlinear system with first order form[39] as follow, $\dot{x} = f(x, t) + u(t)$, and defined control law as $u(t) = \theta_1 e(t) + \theta_2 \int_0^t e(\tau) d\tau$, then we have $f(y^*, t) = f(y^*, 0)$ for any $f \in \mathcal{F}_L$, the closed-loop control system satisfies $\lim_{t \to \infty} x(t) = y^*$ for any initial states $x(0) \in \mathbb{R}$.

## 3 | DESIGN OF UNIVERSAL SELF-LEARNING CONTROL SCHEME

Based on the problem definition in Section 2, we aim to design a controller framework with universal self-learning capabilities. This section introduces the proposed overall controller framework and related details, including the offline training data collection, policy optimization neural network, training methods and pipeline of online self-learning controller.

## 3.1 | Physical performance indicators dataset

Since the system is aimed at more general performance evaluation indicators, we innovatively use the time domain response indicator as the input of the controller update strategy to optimize the neural network. Then, when facing more general control objects, we rely on universal strategies to learn and reconstruct the controller parameters. Here, we define the physical performance index vector as

$$\begin{aligned} d_i &= [\phi_\sigma, \phi_e, \phi_{t_r}, \phi_{t_t}, \Delta K_{\phi 1}, \Delta K_{\phi 2}, \Delta K_{\phi 3}]^\top \\ &= [\phi_\sigma, y^* - y_{end}, t_{rise} - t_0, t_{y^*(1-\Delta_{0.05})} - t_0, \Delta \mathcal{K}]^\top \end{aligned} \quad (2)$$

where $\phi_\sigma = \frac{y(t_{peak}) - y(\infty)}{y(\infty)}$, and $\Delta \mathcal{K} = [\Delta K_{\phi 1}, \Delta K_{\phi 2}, \Delta K_{\phi 3}]$.

More generally, performance evaluation of controllers involves as follows: 1) **Stability**: Assesses whether the control system can maintain or return to a stable state when subjected to external disturbances, as e.g. the overshoot $\phi_\sigma$ of system. 2) **Response speed**: Measures the system's response time to changes in input signals. Common indicators include rise time $\phi_{t_r}$, peak time, and settling time $\phi_{t_t}$. 3) **Steady-state error**: Evaluates the difference between the control system's output and the desired value when the system has reached a steady state, presented as $e_{ss} = y^* - y(\infty)$. Furthermore, there are other performance indicators, such as energy consumption, Integral of Time-weighted Squared Error (ITSE), etc. Build on these evaluation criteria, we propose a model-free neural network for performance optimization. This approach integrates traditional control performance indicators and simulates the human brain's offline optimization processes.

Since our evaluation focuses on the input and output information during the operation of the actual system to assess its response in the environment, there is no need for precise knowledge of the type of control plant. Consequently, a universal policy optimization learning method can be employed. Due to the system noise or uncertain disturbances in real-world engineering systems are unknown, the collected system responses may contain errors and typically have a small sample set. To mitigate this, we extensively augmented the training set data in the simulation environment. Additionally, to prevent the prediction amplitude from exceeding boundaries, we implemented limiting protection when predicting the optimization direction and amplitude.

As shown in Figure 2, the three mutually perpendicular axes represent the basis vectors that determine the control system's performance. This study addresses the simultaneous performance optimization of multiple variables, necessitating a comparison of three different cost calculation methods as equation (3a)-(3c), and here we choose equation (3c).

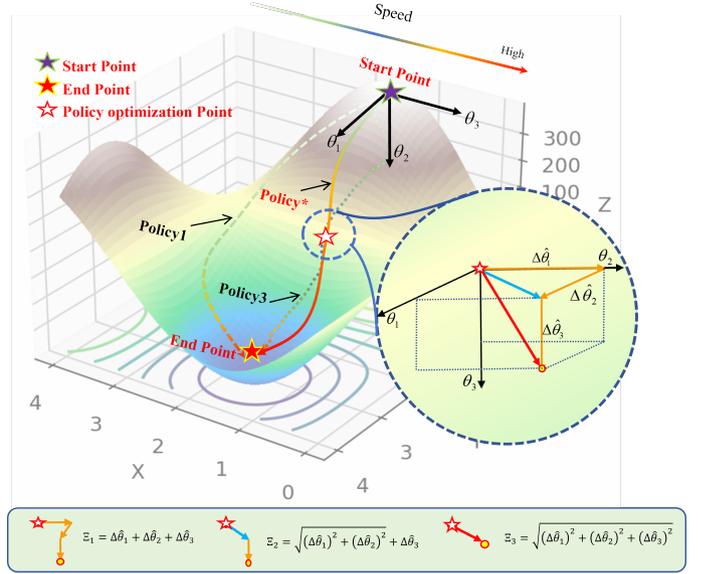

**FIGURE 2** Optimize controller performance to the local minimum point of the surface through different controller parameter adjustment strategies.

$$\Xi_1 = \Delta\hat{\theta}_1 + \Delta\hat{\theta}_2 + \Delta\hat{\theta}_3, \quad (3a)$$

$$\Xi_2 = \sqrt{(\Delta\hat{\theta}_1)^2 + (\Delta\hat{\theta}_2)^2} + \Delta\hat{\theta}_3, \quad (3b)$$

$$\Xi_3 = \sqrt{(\Delta\hat{\theta}_1)^2 + (\Delta\hat{\theta}_2)^2 + (\Delta\hat{\theta}_3)^2}. \quad (3c)$$

*Remark 3.* For each learning cycle $\Delta T$, the three-dimensional coordinate transformation in the small box (in Figure 2) by the selected optimal path to update the controller parameters which could affect the system performance.

*Remark 4.* We designed a cost gradient update algorithm with adjustable weights to compute the multivariable coupling performance cost. We selected the fastest route for path calculation, $\Xi_3$, as presented in Figure 2. This algorithm improves the control system's capability to handle diverse performance metrics across various tasks, each with distinct priorities. For example, in aircraft attitude tracking, the primary goal is to minimize control response time, whereas in surgical robot control, the emphasis is on minimizing overshoot to ensure precision.

## 3.2 | Universal performance policy-optimization learning network

In traditional model-based methods, controllers for fixed-wing UAVs with varying power and wingspans need to be re-modeled. The design and tuning of the controller are then performed using frequency response or other state-space design methods. Alternatively, model-free methods[40] describe the system or its uncertain parts using data from interactions between the model and the environment. Consequently, the effectiveness of the final neural network model primarily depends on the accuracy of data measurement. Moreover, different aircraft and working conditions require fresh data collection and network training, making the process cumbersome and time-consuming. Here, the performance cost assessment based on crucial frequency domain response indicators, this paper introduces an output response performance indicator based on unit step responses to evaluate current controller parameters. Four typical performance indicators are used as inputs to the universal performance optimization policy network. The network outputs an estimate of the future tuning increment of the controller, including direction and amplitude. The universal performance optimization policy network is trained using the dataset described in Section 1. To ensure balanced data distribution among the indicators, distributed normalization parameters are applied before training. The trained policy network iteratively refines the optimization direction and amplitude to approximate the optimal controller parameters. This innovative approach focuses on finding the optimal policy rather than directly identifying the optimal parameters, eliminating the need to address differences between various models. By focusing on the direction towards the true value of the optimal system equilibrium point, the optimal equilibrium point can be found after a limited number of iterations.

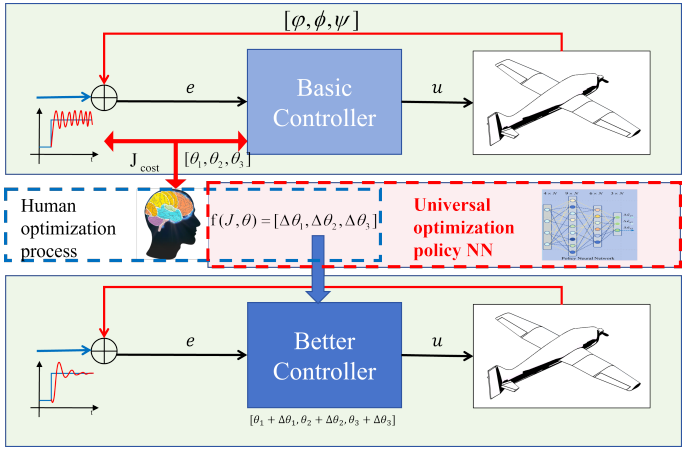

**FIGURE 3** Using aircraft flight controller learning as an example, we initially employ a controller with stable performance derived from historical data of similar models as the baseline. We then apply the proposed USLC method to facilitate in-flight learning, continuously interacting with the environment and dynamically updating the controller in real time until it converges to the desired optimal performance.

## 3.3 | Real-time model-free self-learning controller reconstruction

After completing the training of the universal policy optimization self-learning network, the neural network can be deployed in the target environment. During system operation, periodic step signal excitations are applied in real-time. The performance indicators are calculated and fed into the policy network to predict the controller update direction and amplitude. These predictions are then used to update the controller parameters in real-time, enabling the real-time reconstruction of the controller.

This process is repeated iteratively: step signal responses are periodically introduced, performance indicators are calculated, and the controller is updated until the system's key indicators meet the standards and the overall performance cost is within the predefined threshold. This ensures that the system meets the required performance criteria. The overall performance cost is established to balance the variations caused by different cost preferences in various scenarios. The complete algorithm for the self-learning method for controller policy optimization based on physical performance is presented in Algorithm 1.

In this study, a group of high-performing controllers from the actual system is used as the target reference controllers. The optimal performance response is defined as $[\phi_\sigma, \phi_e, \phi_{t_r}, \phi_{t_t}] = [0.55\%, 0.01, 0.034, 0.021]$. The corresponding best controller parameter set is $[K_1^*, K_2^*, K_3^*, N] = [32.17, 18.6, 12.36, 11320]$. The initial parameters are set to $[a, b, c, d] = [0.6, 0.3, 0.2, 0.3]$, with the number of learning iterations $n = 100$. $K^*$ represents the new controller with updated parameters, and $\mathcal{W}$ denotes

of nonlinear dynamic system was set as

$$\mathcal{J}_{cost}(\theta) = arg \min \sqrt{\mathcal{H}} \quad (4)$$

where performance functions defined as:

$$\mathcal{H} = a(\phi_\sigma^* - \hat{\phi}_{\sigma(k)})^2 + b(\phi_e^* - \hat{\phi}_{e(k)})^2 \\ + c(\phi_{t_r}^* - \hat{\phi}_{t_r(k)})^2 + d(\phi_{t_t}^* - \hat{\phi}_{t_t(k)})^2 \quad (5)$$

with performance wights $a, b, c, d \in \mathbb{R}^+$, especially for tasks that require adjustments to bias specific performance metrics.

Drawing inspiration from flight control design experts, who estimate future controller parameter optimization strategies

**Algorithm 1** USAL via Physical Performance Policy Optimization Learning

```
Data Preprocessing: Let Δ = random[–5%,+5%]
and d_i = (1 + Δ) * d̄_i, then ΔK̂_φ = K*_φ − αK^i_φ,
here, learning rate is α ∈ [0,1], initial it
with α = 0.1.
Input: History performance optimization
policy dataset 𝔻^((4+3)×n), where ith data unit
d_i ∈ 𝔻^((4+3)×n) is defined as Equation (2).
Output: Controller Parameter Change
Estimation [ΔK̂_φ1, ΔK̂_φ2, ΔK̂_φ3]
Initial: Load Policy Neural Network Model M
and system parameters.
Performance Optimization:
for  do i ≤ n , K* ← K(i)
   Step1: Environmental Interaction Response
𝒲 ⇔ K*.
   Step2: Performance Cost Assessment:
(see Equation (4) in Section 3.2)
   if  𝒥_cost< 𝒥* then
       Policy prediction use NN 𝒩:ΔK̂ ⇐ M(i)
       Update Best Policy:ΔK̂(i + 1) ⇐ α * ΔK̂(k)
       K̂(i + 1) ⇐ K_0 + ΔK̂(i + 1)
   end if
   if   d_1(i) < d*_1 or 𝒥_cost(i) < 𝒥* then
       Learned Best Controller:K* ← K
   else
       Keep Learning:  i = i + 1; K(i + 1) ← K̂(i + 1);
K ← K(i + 1);
   end if
end for
```

the unknown disturbances. The system's performance indicators can be obtained by equation (2). The actual network input will combine the $n$ data columns into tensor data for training. The final output of the policy learning network is the estimated controller increment value $\zeta = [a\hat{\beta}_1, b\hat{\beta}_1, c\hat{\beta}_3]^\top$, as illustrated in Figure 1.

## 4 | STABILITY PROOF

Define a class of nonlinear uncertainties system as

$$\begin{cases} \dot{x}_1 = x_2, \\ \dot{x}_2 = f(x_1, x_2, t) + u(t), \end{cases} \quad (6)$$

where $x_1, x_2 \in \mathbb{R}$, and set basic controller as $u(t) = [\theta_1, \theta_2, \theta_3][e(t), \int e(\tau)d\tau, \dot{e}(t)]^\top$, denotes $g(z, y, t) = f(z, y^* - y, t) + f(y^*, 0, t)$. Employed $f = f(x_1, x_2, t)$ represents the uncertain system with known upper bounds that satisfies the Lipschitz continuity condition. The parameters $(\theta_1, \theta_2, \theta_3)$ pertain to the controller structures, e.g., the widely used PID controller structure can defined as $(\theta_1, \theta_2, \theta_3) = (K_p, K_i, K_d)$. We define the space of the function as follows:

$$\mathcal{F}_{L_1,L_2} = \left\{ f \in C_1(\mathbb{R}^2 \times \mathbb{R}^+) \middle| \frac{\partial f}{\partial x_1} \leqslant L_1, \left| \frac{\partial f}{\partial x_2} \right| \leqslant L_2 \right\}, \quad (7)$$

where $\forall x_1, x_2 \in \mathbb{R}$, $\forall t \in \mathbb{R}^+$, and $C_1$ is locally Lipschitz space of function continuous time in $(x_1, x_2)$, and the $L_1, L_2$ are positive constants, the $f$ have continuous partical derivatives with respect to $(x_1, x_2)$.

**Lemma 1.** *Assume* $\Gamma = \frac{\varpi p_0 + \theta_2}{2p_0 + \frac{L_2^2}{2}} > 0$, *then the matrix* **A** *formed as*

$$\mathbf{A} = \begin{bmatrix} 2(\Gamma p(y) - \theta_2) & \Gamma(\varpi + l(\cdot) - \theta_3) \\ \Gamma(\varpi + l(\cdot) - \theta_3) & 2(\Gamma + l(\cdot) - \theta_3) \end{bmatrix} \quad (8)$$

*is a positive definite matrix.*

*Proof of Lemma 1 as Appendix A .* □

Here, we have the stability theorem as:

**Theorem 1.** *For nonlinear uncertain systems (6) with unknown exact models* $f \in \mathcal{F}_{L_1,L_2}$, *with* $f(y, 0, t) = f(y, 0, 0), t \in \mathbb{R}^+, y \in \mathbb{R}, y^* \in \mathbb{R}$. *Then for any* $L_1 > 0, L_2 > 0, \theta_2 > 0, \theta_3 > 0$, *there exists at least one non-empty three-dimensional manifold space* $\Omega_\theta \subset \mathbb{R}^3$, *formed as*

$$\Omega_\theta = \left\{ \Theta \in \mathbb{R}^3 \middle| \theta_1 > L_1, \theta_3 > L_2, \frac{(\theta_1 - L_1)(\theta_3 - L_2) - \theta_2}{\sqrt{\theta_2(\theta_3 + L_2)}} > L_2 \right\}, \quad (9)$$

*such that, for any controller parameters* $\Theta = (\theta_1, \theta_2, \theta_3)$ *in three dimensional manifold* $\Omega_\theta$ *with any initial value* $(x_1(0), x_2(0)) \in \mathbb{R}^2$, *the nonlinear closed-loop system (6) have* $x_1(\infty) \to y^*$ *and* $x_2(\infty) \to 0$.

*Proof of Theorem 1.* Since $\dot{x} = f(x, t) + u(t)$ as defined in equation (6), we denote

$$\begin{cases} x(t) = \int_0^t e(s)\, ds + \frac{f(y^*, 0, 0)}{\theta_2}, \\ y(t) = e(t), \\ z(t) = \dot{e}(t), \\ g(y, z, t) = -f((y^* - y), -z, t) + f(y^*, 0, t), \end{cases} \quad (10)$$

then, we can rewritten equation (6) as

$$\begin{cases} \dot{x} = y, \\ \dot{y} = z, \\ \dot{z} = g(y, z, t) - \theta_1 y - \theta_2 x - \theta_3 z. \end{cases} \quad (11)$$

For any $t \in [0, +\infty)$ and $f \in \mathcal{F}_{L1,L2}$, it is easy to obtained that $g \in \mathcal{F}_{L1,L2}$ and the initial point $g(0, 0, t) = 0$. Further we rewrite the formula as $g(y, z, t) = h(y, t)y + l(y, z, t)z$.

Here, we define the part as

$$h(y,t) = \begin{cases} \dfrac{g(y,0,t)}{y}, & y \neq 0, \\ \dfrac{\partial g}{\partial y}(0,0,t), & y = 0, \end{cases} \quad (12)$$

and

$$l(y,z,t) = \begin{cases} \dfrac{g(y,z,t) - g(y,0,t)}{z}, & z \neq 0, \\ \dfrac{\partial g}{\partial z}(y,0,t), & z = 0. \end{cases} \quad (13)$$

Since $t \geq 0, y \in \mathbb{R}, f(y,0,t) = f(y,0,0,)$, combined with (12) and (13), we have $h(y,t) \leqslant L_1$, $|l(y,z,t)| \leqslant L_2$. By the mean value theorem and (6), hence $h(y,t) = \frac{1}{y}g(y,0,0)$, and $h(y,t)$ could represented as $h(y)$.

Rewritten the system 11 as

$$\dot{\mathbf{X}} = \mathbf{Q}(X,t)\mathbf{X} \quad (14)$$

where $\mathbf{X} = [x,y,z]^\top$ and

$$\mathbf{Q}(x,y,z,t) = \begin{bmatrix} 0 & 1 & 0 \\ 0 & 0 & 1 \\ -\theta_2 & -\theta_1 + h(y) & -\theta_3 + l(y,z,t) \end{bmatrix}, \quad (15)$$

such that the equation (11) can be rewritten as

$$[\dot{x},\dot{y},\dot{z}]^\top = \mathbf{Q}(\cdot)[x,y,z]^\top. \quad (16)$$

Construct a Lyapunov function as follow

$$V(x,y,z) = [x,y,z]\mathbf{M}[x,y,z]^\top + \int_0^y (p(\tau) - p_0)\tau \mathrm{d}\tau, \quad (17)$$

where $\varpi = \frac{1}{2}(\varpi_0 + \varpi_1)$, and

$$\begin{cases} \varpi_0 = \inf\{\theta_3 - l(y,z,t)\}, \\ \varpi_1 = \sup\{\theta_3 - l(y,z,t)\}, \\ p_0 = \inf\{p(y)\}, \\ p(y) = \theta_1 - h(y). \end{cases} \quad (18)$$

By the fact that $\theta_1 > L_1, \theta_3 > L_2$, it is easy to get that $\varpi_0 \geq \theta_3 - L_2 > 0, p_0 \geq \theta_1 - L_1 > 0$, here matrix $\mathbf{M}$ represents

$$\mathbf{M} = \frac{1}{2}\begin{bmatrix} \Gamma\theta_2 & \theta_2 & 0 \\ \theta_2 & p_0 + \Gamma\varpi & \Gamma \\ 0 & \Gamma & 1 \end{bmatrix}, \text{with } \Gamma = \frac{\varpi_0 p_0 + \theta_2}{2p_0 + \frac{L_2^2}{2}} > 0. \quad (19)$$

We assume that matrix $\mathbf{M}$ is a positive definite matrix. We proofed it in Appendix B. It is not difficult to see that $V$ is an unbounded positive definite function in three-dimensional space. Therefore, the derivative of the Lyapunov function as equation (17) can be obtained

$$\begin{aligned}\dot{V}(x,y,z) &= [x,y,z](\mathbf{M}Q(x,y,z,t) + Q^\top(x,y,z,t)\mathbf{M})[x,y,z]^\top \\ &\quad + (p(y) - p_0)yz \\ &= [y,z]\begin{bmatrix} \theta_2 - \Gamma p(y) & \frac{\Gamma(\varpi + l(\cdot) - \theta_3) + p_0 - p(y)}{2} \\ \frac{\Gamma(\varpi + l(\cdot) - \theta_3) + p_0 - p(y)}{2} & \Gamma + l(\cdot) - \theta_3 \end{bmatrix}\begin{bmatrix} y \\ z \end{bmatrix} \\ &\quad + [p(y), -p_0]\begin{bmatrix} y \\ z \end{bmatrix} \\ &= [y,z]\begin{bmatrix} \theta_2 - \Gamma p(y) & \frac{\Gamma(\varpi + l(\cdot) - \theta_3)}{2} \\ \frac{\Gamma(\varpi + l(\cdot) - \theta_3)}{2} & \Gamma + l(\cdot) - \theta_3 \end{bmatrix}\begin{bmatrix} y \\ z \end{bmatrix} \\ &= -[y,z]\mathbf{P}(y,z,t)\begin{bmatrix} y \\ z \end{bmatrix}, \end{aligned} \quad (20)$$

where

$$\mathbf{P} = \frac{1}{2}\mathbf{A} = -\begin{bmatrix} \theta_2 - \Gamma p(y) & \frac{\Gamma(\varpi + l(\cdot) - \theta_3)}{2} \\ \frac{\Gamma(\varpi + l(\cdot) - \theta_3)}{2} & \Gamma + l(\cdot) - \theta_3 \end{bmatrix}. \quad (21)$$

As mentioned in Lemma 1, $\mathbf{P}$ denotes a symmetric matrix, and we proofed that it is actually positive definite for any $t \in \mathbb{R}^+$, and $y,z \in \mathbb{R}$. The proofs details can find in Appendix A.

We can denote the minimum eigenvalue of $P(\cdot)$ as

$$\begin{aligned}\lambda_{\min}\{\mathbf{P}(y,z,t)\} &= \mathcal{K}(y,\gamma,\delta) \\ &\triangleq \frac{1}{2}\{\Gamma p(y) - \theta_2 + \delta - \sigma\},\end{aligned} \quad (22)$$

where $\sigma = \sqrt{(\Gamma p(y) - \theta_2 - \delta)^2 + 4\gamma^2}$ and $\lambda(y) = \inf[\mathcal{K}(y,\gamma,\delta)]$ denotes the infimum for any $\gamma,\delta$ satisfied

$$|\gamma| \leqslant \frac{\Gamma L_2}{2}, \quad (23a)$$

$$\varpi_0 - \Gamma \leq \delta \leq \varpi_1 - \Gamma_0. \quad (23b)$$

For the matrix $\mathbf{P}$ is positive definite, then the eigenvalue function $\lambda(y)$ is a positive continuous function of $y$. Hence the solution of the system (6) satisfy

$$\lambda(y(t))(y(t)^2 + z(t)^2) \to 0, \quad (24)$$

when $t \to \infty$, we can get $\lambda(y) > \epsilon > 0$. Thus, $\lim\limits_{t \to \infty}(y(t)^2 + z(t)^2) = 0$, that is $x_1(\infty) \to y^*$ and $x_2(\infty) \to 0$.
Then, the proof of Theorem 1 is completed. □

## 5 | SIMULATION

In this section, numerical simulations are designed for an aircraft attitude control task[27,36] based on stochastic aerodynamic model deviation to verify the effect of the algorithm on global stable convergence, control response accuracy, and response speed. At the same time, traditional model-based adaptive method[5] are utilized to compare the effects with our approach.

## 5.1 | Case A: Numerical simulation of low-order circuit control systems

The aim of the numerical simulation is to verify that the proposed method can reconstruct the optimal controller through a finite number of self-learning iterations, even with initial parameter variations in a conventional second-order circuit system. Additionally, the proposed method is compared to time-domain adaptive tuning control based on a traditional precise model. A simplified second-order model of the circuit system is established as follows:

$$G_1(s) = \frac{c_1}{s^2 + c_2 s + c_3}, \quad (25)$$

where $c_1 = 8, c_2 = 0.878, c_3 = 21.5$, and the Proportional–Integral–Derivative (PID) controller structure is applied in unit negative feedback control system, the initial gains in the controller were set as $\theta_{kp} = 90, \theta_{ki} = 3, \theta_{kd} = 1, N = 100$. Next, the control effect is evaluated in real time based on the unit step output response of the system.

### 5.1.1 | a) Validity verification

The controller structure parameters are continuously updated and gradually and quickly iterated to the optimal performance. When the controller reaches the optimal performance is considered as the optimal controller, and the corresponding parameters are the best controller parameters. It can be viewed from Figure 4, the overshoot of the optimal controller for random second-order system performance is only 0.2%, the rise time is 0.015*s*, and the stabilization time is 0.03*s*. At the same time, it can be seen from the figure that the maximum overshoot during the learning process is 8%. Therefore, the performance improvement of our learned controller is very significant.

The prediction curve shows that after the controller has some fluctuations in the early optimization stage, the multi-variable parameters will gradually converge to a smaller interval simultaneously, which is consistent with approaching optimal control. When the controller runs, the optimization quantity gradually approaches 0 in the stability theory.

### 5.1.2 | b) Policy learning initial value robustness verification

The above experiments verified the effectiveness of the USLC algorithm in numerical simulation can be seen as following figures: Figure 8-Figure 11.

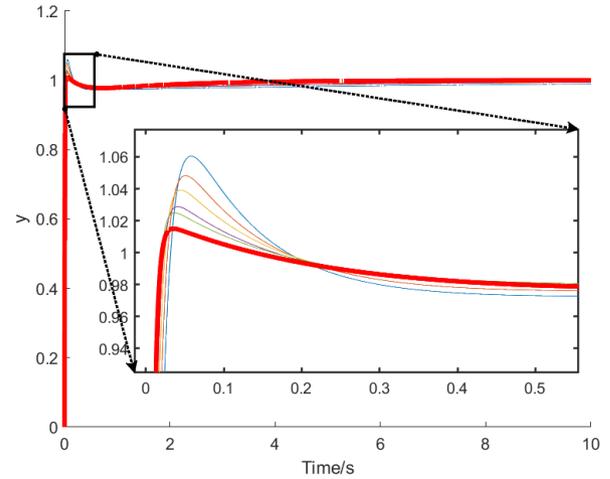

**FIGURE 4** The adaptive learning optimization process of controller parameters based on physical performance policy for $G_1$ starts with the initial control system parameters set to $\theta_0 = [90, 10, 10]$. The thick red line represents the control response of the optimized controller that meets the performance indicators. The thin lines in different colors show the controller response curves during the learning process. The reference command of the designed controller is a step input with an amplitude of 1 at $t = 0$s.

As Figure 8 shows, the control system's initial parameters $G_1$ are set to $\theta_0 = [10, 1, 1]$. The thick red line is the control response corresponding to the learned controller that meets the performance indicators. The thin lines in different colors are the controller response curves during the learning process, and the controller target is the step instruction with an amplitude of 1 at $t_0 = 0$s. It can be seen that the response overshoot of the optimal controller of the system is about 0.3%, the rise time is 0.005s, the stabilization time is 1.2s, and the maximum overshoot during the learning process is about 11%. As shown in Figure 9, the controller optimization estimator begins to level off.

Comparing Figure 4 and Figure 5 in subsection 5.1.1, we can see that when the initial value of the controller is randomly initialized, the estimator of the universal self-learning process will fluctuate relatively large. The learning time will become more prolonged, but it can still result in better performance. Experimental results show that the initial value robustness of the proposed method is good.

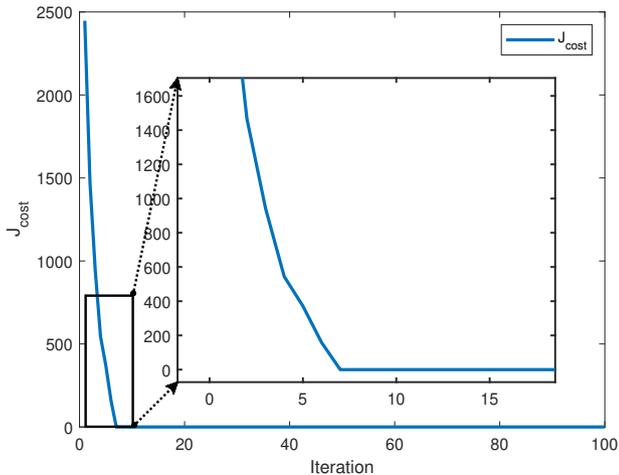

**FIGURE 5** The real-time system control $G_1$ performance cost curve during the controller's real-time learning and reconstruction process quickly converges to a stable state, indicating that the system is gradually approaching the global optimal solution.

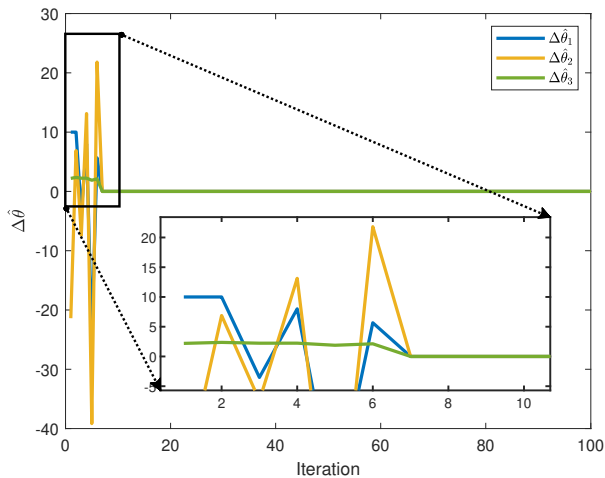

**FIGURE 6** Based on the physical performance policy optimization neural network, prediction of the real-time optimization amount of the controller for $G_1$.

## 5.2 | Case B: Simulation of high-order morphing fixed-wing flight control

After completing the primary numerical verification, we designed a verification case in a variable-span fixed-wing aircraft attitude control; we designed and implemented an online adaptation of the attitude controller with an unknown dynamic system model. We selected a nonlinear model in the literature

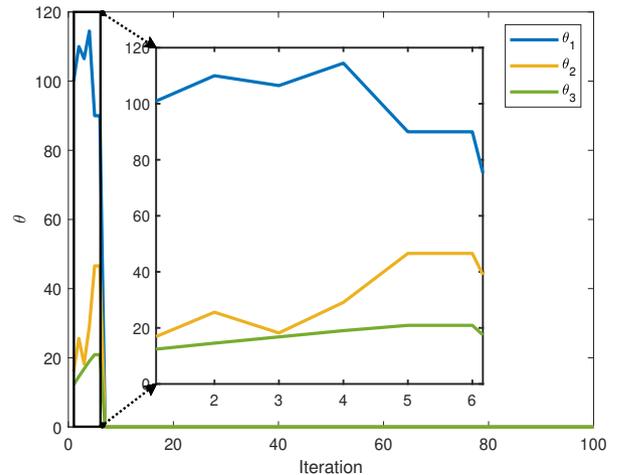

**FIGURE 7** Real-time updated $G_1$'s controller structural parameter change curve.

for learning control with a random model parameter bias of 20%.

This case focuses on the attitude flight control task of fixed-wing UAVs. It is well-known that the attitude flight control of fixed-wing UAVs exhibits a strong coupling relationship among the control channels. To simplify the controller design and implementation, these three control channels are typically decoupled for analysis and design, and their control distributions are then summed. To more intuitively verify our proposed method, this paper examines explicitly the roll channel's attitude control after decoupling. Here, $x_1 = \phi$ represents the roll angle, and $x_2 = \dot{\phi}$ represents the roll rate. Real-time sensor data is generally obtained from the flight control Attitude and Heading Reference System (AHRS) module or a visual motion capture system.

We selects two states of fixed-wing wingspan morphing, as shown in Figure 12, to verify USAL in the simulation environment. Typically, the pitch angle attitude control transfer function of a fixed-wing UAV can be defined as follows:

$$G_2(s) = \frac{\bar{c}_1 s^2 + \bar{c}_2 s + \bar{c}_3}{\bar{c}_4 s^4 + \bar{c}_5 s^3 + \bar{c}_6 s^2 + \bar{c}_7 s + \bar{c}_8} \quad (26)$$

Based on this transfer function structure, as shown in the following Table 1, we construct two sets of longitudinal attitude control transfer functions with different parameters to verify the effectiveness of the self-learning algorithm when the wingspan of the aircraft changes.

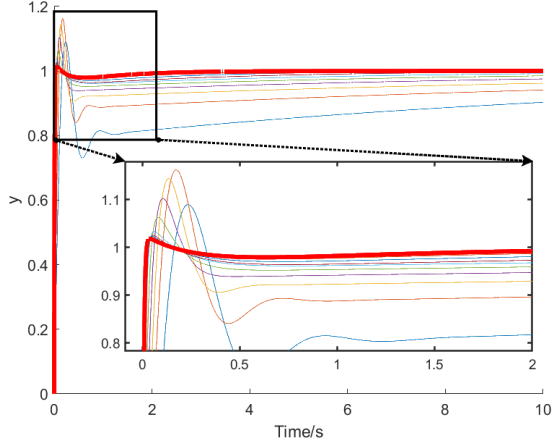

**FIGURE 8** Self-learning optimization process of controller parameters based on physical performance optimized policy network as present in Figure 1.

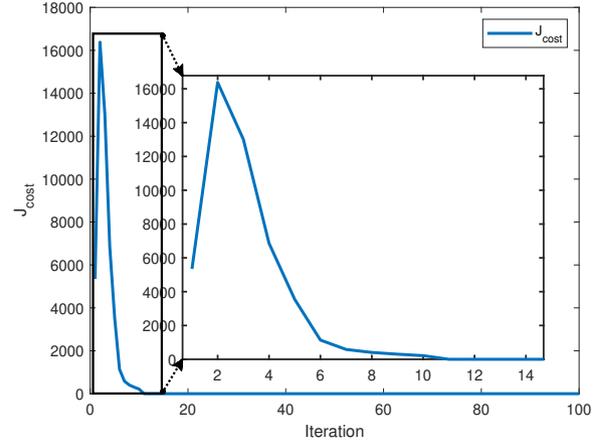

**FIGURE 9** For $G_1$, performance cost curve of self-learning controller with random initial value. After the $9th$ learning, the time that can stand by tends to the neighborhood near 0.

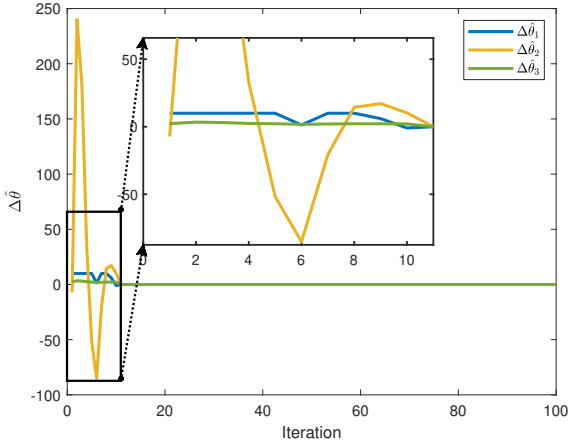

**FIGURE 10** Based on the physical performance policy optimization neural network, for controller $G_1$ optimization estimation in the case of random initial values.

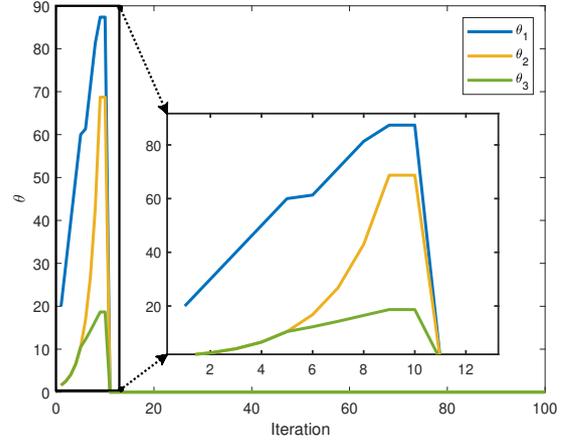

**FIGURE 11** In the case of random initialization of controller parameter initial values, real-time updated $G_1$'s controller structural parameter change curve.

### 5.2.1 | a) Wingspan A: $G_{2A}$

We do not need to identify the aircraft model in advance for the two simulated flight states. We use the same USLC framework and interact with the environment during flight. Then, we iterate the controller parameters in each learning cycle and finally learn the controller parameters that meet the preset performance indicators. The two learning processes are entirely independent, which proves that this method has good versatility in aircraft of different modes, as shown in Figure 13-16. The response results depicted in Figure 13 demonstrate the proposed method's efficacy in handling the high-order model's fixed wing. The performance cost change curve presented in Figure 14 illustrates the iterative self-learning process of the controller. Comparatively, in contrast with low-order systems, the proposed method effectively diminishes the cost of physical performance error. However, it is noteworthy that the learning rate may correspondingly decrease.

### 5.2.2 | b) Wingspan B: $G_{2B}$

As mentioned in Figure 12, when the aircraft needs greater lift during takeoff and landing, the wings need to be morphing to the state B $G_{2B}$, thereby reducing the takeoff and landing distance. Utilizing USLC algorithms and initial parameters, real-time online learning of the flight controller's parameters, and this method is no need to know the model in advance. The corresponding outcomes are depicted in Figure 17-20.

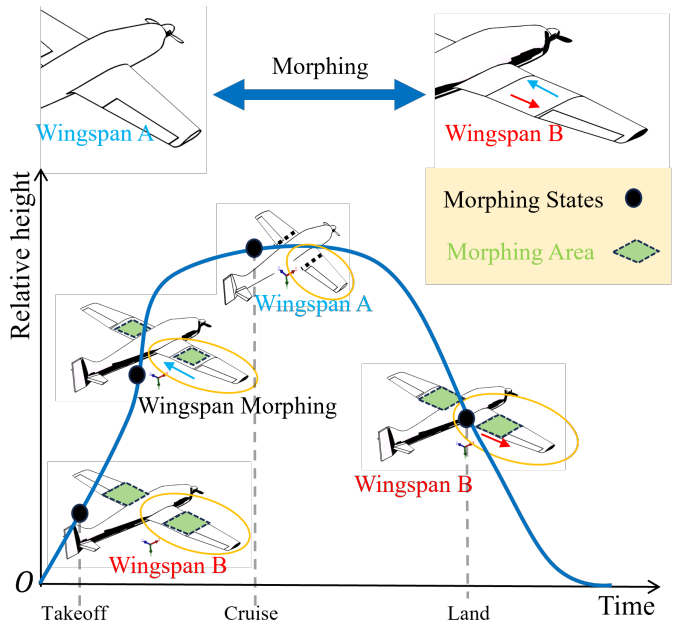

**FIGURE 12** The complete wing morphing process: from takeoff state (Wingspan B) to cruise state (Wingspan A), then land state (Wingspan B). Different aircraft aerodynamic layouts determine the aircraft's flight envelope and flight performance.

**TABLE 1** Variable wingspan aircraft state parameters

| Parameters | Wingspan Status $G_{2A}$ | Wingspan Status $G_{2B}$ |
| --- | --- | --- |
| $\bar{c}_1$ | 0.069 | 0.8 |
| $\bar{c}_2$ | 8.41 | 9.40 |
| $\bar{c}_3$ | 0.91 | 1.04 |
| $\bar{c}_4$ | 1 | 1 |
| $\bar{c}_5$ | 25.14 | 25.01 |
| $\bar{c}_6$ | 161.8 | 159.8 |
| $\bar{c}_7$ | 16.75 | 17.17 |
| $\bar{c}_8$ | 1.28 | -0.46 |

Note: In order to facilitate verification, this paper uses the transfer function simplification method of minor disturbance linear expansion at the equilibrium point to obtain the control object model.

As can be seen from Figure 17, when the plant model of the controller system changes, the self-learning controller still achieves an excellent performance of 2% overshoot even after a period of learning without any extra work. Figure 18 shows that the proposed method can stably converge to the optimal performance region even after the aircraft model undergoes a significant change. Compared the time to system performance reached steady state in Figure 18 and Figure 14, it can be seen that the model of the aircraft itself affects the time to converge to a stable state. The more maneuverable the aircraft, the shorter the learning time.

By comparing Figures 19 and 20 with Figures 15 and 16 in state Wingspan A as mentioned in subsection 5.2.1, we observe that the predicted controller reconstruction deviation trends vary across different control objects. This variation is based on the proposed physical performance strategy optimization network. Nonetheless, these trends generally converge towards optimal performance metrics, aligning with theoretical predictions regarding system stability and response time. Furthermore, Figures 19 and 20 indicate that during the self-learning reconstruction process, the controller may experience significant performance fluctuations due to initial parameter variability.

Therefore, for future deployment on real aircraft, it is crucial to integrate robust safety protection mechanisms, such as fault-tolerant control algorithms, and appropriate limiting filters, to mitigate performance fluctuations and ensure operational safety. This underscores a key contribution of simulation experiments: providing critical insights and safeguards necessary for the safe and effective transition to real-world experiments.

## 6 | CONCLUSIONS

This study delves into the optimization of universal self-learning capabilities of controllers in nonlinear dynamical systems with uncertain models. We introduce an evaluation method using a participant-critic approach in the controller learning process to achieve global convergence with unknown performance objectives. A real-time controller parameter prediction neural network model is developed to generate optimization strategies based on the physical property tensor in the frequency domain step response. The reconfigured controller progressively reaches a steady state as the data capturing the interaction between the object and the environment is updated. The proposed method employs an online policy iterative learning architecture, grounded in the stability theory of nonlinear systems, to ensure stability and adaptability to large scales. Finally, the proposed algorithm is validated on a low-order circuit system and a high-order morphing fixed-wing system. The results demonstrate that the proposed method exhibits good stability and robustness. Future research will extende USLC to address more universal self-learning control problems across a broader range of similar models, such as for various uncrewed vehicles, boats, and aircraft.

**AUTHOR CONTRIBUTIONS**
Yanhui Zhang conceived and designed the study, performed experiments, analyzed data, and wrote the paper. XXX and XXX contributed to critically revised the manuscript. Weifang Chen provided resources and supervised the study. XXX and XXX participated in the suggestion to revise the manuscript. All authors have read and approved the final manuscript.

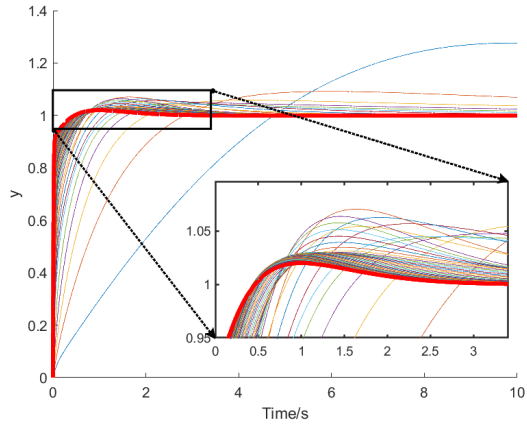

**FIGURE 13** Unit step response effect during the learning process of the lateral flight controller of a UAV with wingspan state A.

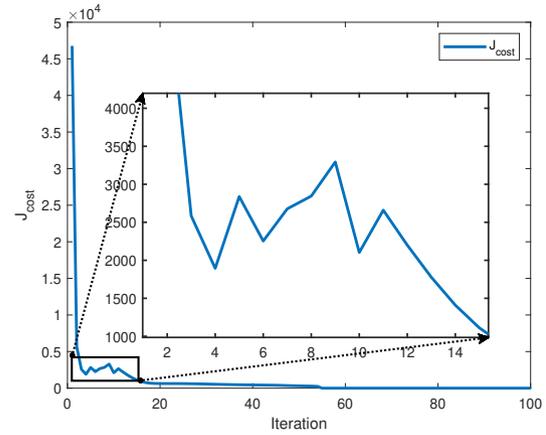

**FIGURE 14** Performance cost change curve of controller iterative self-learning process.

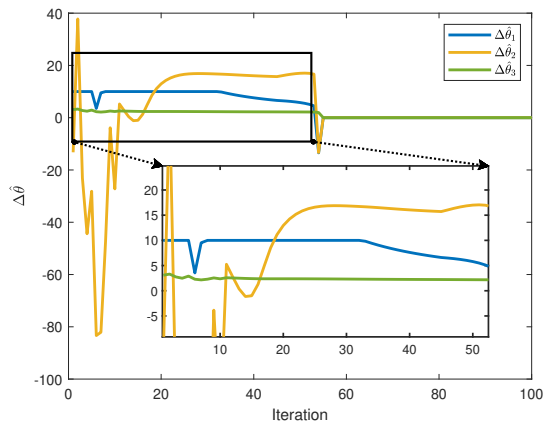

**FIGURE 15** Prediction of controller updates by neural networks for physical performance optimization strategies in universal self-learning processes.

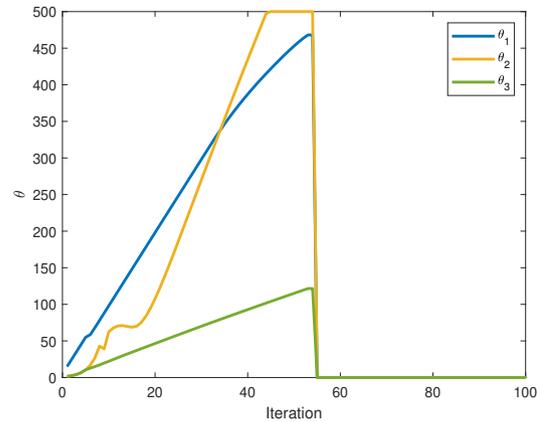

**FIGURE 16** The cumulative change curve of the self-learning controller parameters during the learning process.

## FINANCIAL DISCLOSURE
## CONFLICT OF INTEREST

The authors declare no potential conflict of interests.

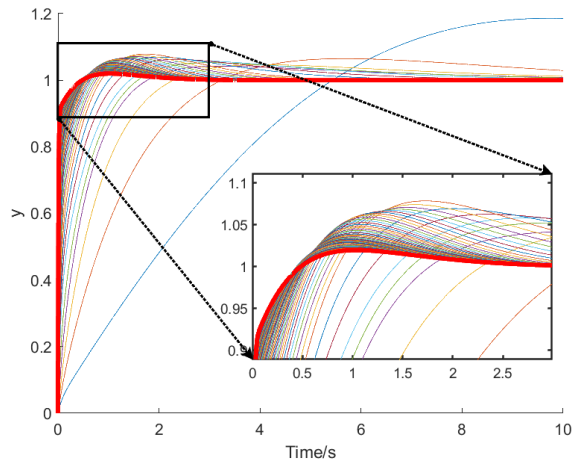

**FIGURE 17** Unit step response effect during the learning process of the lateral flight controller of a UAV with wingspan state B.

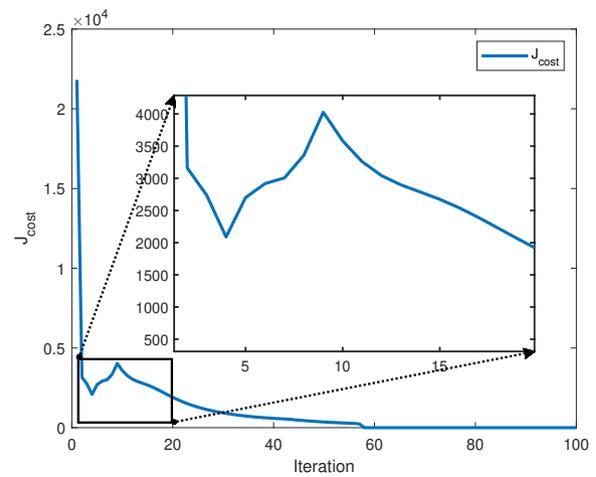

**FIGURE 18** Performance cost change curve of controller iterative self-learning process.

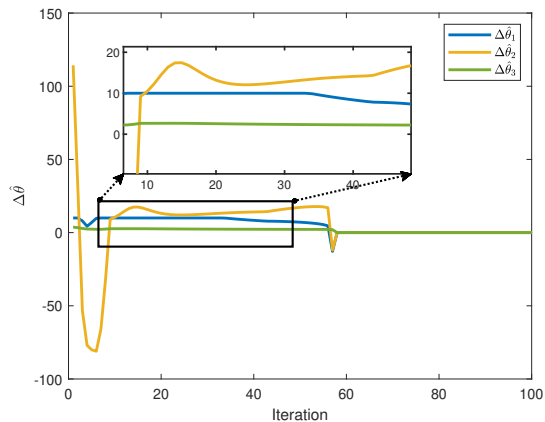

**FIGURE 19** Prediction of controller updates by neural networks for physical performance optimization strategies in universal self-learning processes.

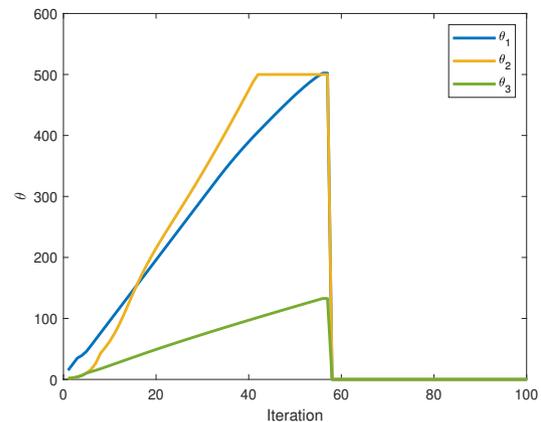

**FIGURE 20** The cumulative change curve of the self-learning controller parameters during the learning process.

**APPENDIX**

## A  PROOF OF LEMMA 1.

*Proof.* Define

$$\begin{cases} \delta = \theta_3 - \Gamma - l(y,z,t), \\ \gamma = \dfrac{\Gamma}{2}[\theta_3 - \varpi - l(y,z,t)], \\ \varpi = \dfrac{1}{2}(\varpi_0 + \varpi_1). \end{cases} \quad (A1)$$

Combined $L_2 \geq |l(\cdot)|$ with the equation (B4) in Appendix B, we get

$$\begin{aligned} \Gamma p(y) - \theta_2 &\geqslant \Gamma p_0 - \theta_2 > 0, \\ \varpi_1 - \Gamma &\geqslant \theta_3 - \Gamma - l(y,z,t) \geqslant \varpi_0 - \Gamma > 0, \\ |\varpi + l(y,z,t) - \theta_3| &\leqslant L_2. \end{aligned} \quad (A2)$$

Therefore, with (A1), (A2) and (B4), we can obtained that

$$\begin{aligned} (\Gamma \varpi(y) - \theta_2)(\theta_3 - \Gamma - l(y,z,t)) &\geqslant (\Gamma p_0 - \theta - 2)(\varpi_0 - \Gamma) \\ &> \dfrac{\Gamma^2 L_2^2}{4} \\ &\geqslant \gamma^2 \geqslant 0. \end{aligned} \quad (A3)$$

Hence, the matix **P** and **A** are proofed as positive definite in $\Omega_\theta$. □

## B  PROOF OF LEMMA 2.

*Proof.* In order to proof the constants matrix **M** is positive definite matrix, we first should proof the follow inequalities hold,

$$\begin{cases} \varpi_0 - \Gamma > 0, \\ \Gamma p_0 - \theta_2 > 0, \\ 4(\Gamma \varpi_0 - \theta_2)(\varpi_0 - \Gamma) - \Gamma^2 L_2^2 > 0, \end{cases} \quad (B4)$$

as $\Omega_\theta$ defined in equation (9), it is easy to obtain

$$(\theta_1 - L_1)(\theta_3 - L_2) > \theta_2 + L_2\sqrt{\theta_2\theta_3 + \theta_2 L_2}, \quad (B5)$$

then, we have $\varpi_0 p_0 > \theta_2 + L_2\sqrt{\theta_2\theta_3 + \theta_2 L_2}$, thus with $\varpi_0 \in (0, \theta_3 + L_2]$, it is not to difficult get

$$(\varpi_0 p_0 - \theta_2)^2 > \theta_2 \varpi_0 L_2^2 \quad (B6)$$

thus

$$\begin{aligned} \Gamma - \varpi_0 &= \dfrac{\varpi_0 p_0 + \theta_2}{2p_0 + \dfrac{L_2^2}{2}} - \varpi_0 \\ &= \dfrac{-(2\varpi_0 p_0 - 2\theta_2 + \varpi_0 L_2^2)}{4\varpi_0 + L_2^2} \\ &< 0 \end{aligned} \quad (B7)$$

That is the first equation in (B4) is true, combined with equation (B6), can have

$$\begin{aligned}
&4(\Gamma\varpi_0 - \theta_2)(\varpi_0 - \Gamma) - \Gamma^2 L_2^2 \\
&= 4(\varpi_0 p_0 + \theta_2)\Gamma - (4\varpi_0 + L_2^2)\Gamma^2 - 4\theta_2\varpi_0 \\
&= \frac{4(\varpi_0 p_0 + \theta_2)^2}{4\varpi_0 + L_2^2} - 4\theta_2\varpi_0 \\
&= \frac{4[(\varpi_0 p_0 + \theta_2)^2 - (4p_0 + L_2^2)\theta_2\varpi_0]}{4\varpi_0 + L_2^2} \\
&= \frac{4[(\varpi_0 p_0 - \theta_2)^2 + 4\varpi_0 p_0 \theta_2 - (4p_0 + L_2^2)\theta_2\varpi_0]}{4\varpi_0 + L_2^2} \\
&= \frac{4[(\varpi_0 p_0 - \theta_2)^2 - L_2^2 \theta_2\varpi_0]}{4\varpi_0 + L_2^2} > 0.
\end{aligned} \quad (B8)$$

As a result, the last equation in (B4) is valid, we known that $\Gamma > 0, \Gamma\theta_2 > 0$ and $\Gamma\varpi_0 > 0$, then

$$\begin{aligned}
\det \begin{bmatrix} \Gamma\theta_2 & \theta_2 \\ \theta_2 & p_0 + \Gamma\varpi \end{bmatrix} &= \Gamma\theta_2(p_0 + \Gamma\varpi) - \theta_2^2 \\
&\geqslant \theta_2(\Gamma p_0 - \theta_2 + \Gamma^2\varpi_0) \\
&> \theta_2 \Gamma^2 \varpi_0 \\
&= \Gamma\theta_2 * \Gamma\varpi_0 \\
&> 0.
\end{aligned} \quad (B9)$$

It follows from the same principle in

$$\begin{aligned}
\det \begin{bmatrix} \Gamma\theta_2 & \theta_2 & 0 \\ \theta_2 & p_0 + \Gamma\varpi & \Gamma \\ 0 & \Gamma & 1 \end{bmatrix} &= \theta_2(\Gamma p_0 - \theta_2 + \Gamma^2\varpi - \Gamma^3) \\
&> \theta_2(\Gamma^2 \varpi_0 - \Gamma^3) \\
&= \theta_2\Gamma^2 * (\varpi_0 - \Gamma) > 0.
\end{aligned} \quad (B10)$$

Hence, the second equation of (B4) is valid, the constant matrix **M** is positive definite. □